\documentclass{IET-Conf-Paper}

\usepackage{tikz}
\usetikzlibrary{arrows.meta,positioning}

\begin{document}

\title{Characterisation of the DQsym Framework Against Harmonic State-Space and Dynamic Phasor Models for Modular Multilevel Converters}

\author{Saif Alsarayreh\ad{1}\corr, Robert Dimitrovski\ad{1}\ad{2}, Aleksandra Leki\'c\ad{1}}

\address{\add{1}{Department of Electrical Sustainable Energy, Delft University of Technology, Delft, The Netherlands}
\add{2}{TenneT TSO GmbH, Bernecker Str. 70, 95448 Bayreuth, Germany}
\email{S.T.S.Alsarayreh@tudelft.nl}}


\begin{abstract}
\label{sec:0_abstract}
Modern power systems rely on power converters that exhibit multifrequency periodic dynamics, such as modular multilevel converters (MMCs). Such converters are commonly analysed using harmonic state-space (HSS) and dynamic phasor (DP) representations. Both approaches rely on building state matrices based on the system topology, whose size grows with the number of harmonics retained in the model. This paper shows that a symmetrical direct-quadrature (dq) dynamic phasor formulation for time-domain simulation, named DQsym, has a smaller solver state matrix than HSS and DP, and whose size is independent of the truncation order $K$. This reduction results from evaluating harmonic coupling outside the state matrix, in the nonlinear signal blocks. The open-loop MMC model used in this paper is derived in three harmonic-domain (HD) formulations and validated against a three-phase time-domain reference model. The models are compared in terms of state-space size and sparsity, harmonic content, steady-state accuracy and eigenvalues. For truncation at $K = 4$, the DQsym solver has 45 states against 135 for HSS and DP, stays within about 4\% of the reference, and, once closed with its coupling blocks, yields eigenvalues identical to those of HSS and DP.


\end{abstract}
\maketitle
\section{Introduction}
\label{sec:introduction}

For power systems and power electronics analysis, engineers rely on Harmonic-Domain (HD) state-space models to expose frequency coupling, such as circulating current harmonics, that dictates interactions between a power converter's internal dynamics and the AC/DC networks it connects to.

HD modelling represents time-varying or linear time-periodic (LTP) electrical systems through the complex Fourier coefficients of their signals, referred to as harmonic vectors in harmonic state-space (HSS) modelling and as dynamic phasors (DP) in generalised averaging~\cite{ghimire2026review,Xu25}. Instead of tracking rapidly oscillating instantaneous waveforms, HD models describe the dynamics of these coefficients, which vary slowly compared with the fundamental period. In DP, each coefficient is computed over a sliding window of one fundamental period, so it is a time-varying quantity~\cite{sanders1991generalized}. In HSS, the signal is expanded with exponentially modulated periodic components $e^{(s + jh\omega_0)t}$ around a periodic steady state, so the coefficients are defined in the Laplace domain~\cite{derua2020comparative,kwon2017piecewise}.

While both methods are frequency-lifted modelling techniques, based on generalised Fourier expansion, they differ in origin and primary focus~\cite{blin_comparison_2021}. HSS stems from the harmonic analysis framework that is structured around block Toeplitz matrices; on the other hand, DP originates from generalised averaging concepts with the primary focus on increasing the simulation time step to simulate large interconnected power system networks efficiently~\cite{sanders1991generalized,n12}. 

HSS relies on a block-banded, block-Toeplitz LTI state-space representation over a contiguous harmonic range, $h \in \{-K,\dots,K\}$, thereby converting LTP power converters, such as a Modular Multilevel Converter (MMC), into an equivalent LTI state-space model. HSS is widely used for small-signal stability analysis and exact harmonic impedance extraction~\cite{kwon2017piecewise,Becker24}. On the other hand, the DP approach retains only non-negative harmonic orders $h \in \{0,\dots, K\}$ and is typically implemented in a real-valued form to reduce system computational burden~\cite{jamshidifar2016small,nwaneto_equivalent_2025,nwaneto_full-order_2023,choi_dynamic_2017}.
Analytically, both approaches share a structural cost: the state space matrix size scales with the truncation order $K$, thereby affecting the state matrix sparsity and the resulting harmonic coupling. In addition, the sparsity and coupling structure only becomes apparent after the full harmonic-domain matrices are derived.

To avoid this dependence of the state matrix on the number of harmonic orders, the DQsym framework~\cite{alsarayreh_dynamic_2027,SaifOAPJE,ALSARAYREH2026102719} represents harmonics in the signals rather than in the states. Each signal is expressed as a $3\times(K+1)$ complex matrix, whose rows correspond to the positive-, negative-, and zero-sequence components and whose columns correspond to the DC term and harmonic orders $1$ to $K$, with each entry given in $d+jq$ form. The system matrices of the studied model are therefore retained without augmentation, regardless of $K$. Harmonic coupling is introduced only where signals are multiplied, in coupling blocks outside the solver, so the coupling structure follows directly from the block topology rather than from a derived global matrix. The computational cost is thereby shifted from the size of the solver state matrix to the width of the signals and the convolutions evaluated at each time step.

Throughout this paper, DQsym solver refers to the arm- and neutral-current dynamics integrated by the solver, and DQsym to the complete model including the capacitor and multiplication blocks. 
The contributions of this paper are as follows:
\begin{itemize}
    \item Derivation of an open-loop MMC model in the $abc$ frame and in three harmonic-domain formulations: HSS, DP, and DQsym.
    \item A comparison of the three formulations for the same MMC at a matched harmonic truncation order, in terms of state-vector size, state-matrix sparsity, harmonic content, time-domain accuracy against the $abc$ reference, and eigenvalues.
    \item A demonstration that DQsym achieves accuracy comparable to HSS and DP with a diagonal, $K$-independent solver state matrix and that closing the coupling blocks around the solver recovers the same eigenvalues.
\end{itemize}

The remainder of this paper is organised as follows. Section~\ref{sec:harmonic_formulation} reviews the HSS and DP formulations and outlines the mathematical approach adopted in this work. Section~\ref{sec:model} develops the MMC model in the $abc$ frame and derives the corresponding HD models from the converter's ordinary differential equations. Section~\ref{sec:resluts} presents and discusses the results of the analysis. Finally, Section~\ref{sec:conclusion} concludes the paper.

\section{Harmonic Domain Formulations}
\label{sec:harmonic_formulation}
Building a model in HD relies on three fundamental mathematical operations:
\begin{itemize}
    \item Fourier Series Expansion: A time-domain signal $x(t)$ is expanded into Fourier components using
\begin{equation}
    x(t) \approx \sum_{h=-K}^{K} \langle x\rangle_h(t)\, e^{jh\omega_0 t}
\end{equation}
where $\langle x \rangle_h(t)$ is the $h$-th dynamic phasor or harmonic, and $h \in \mathbb{Z}$ is the harmonic order, corresponding to the frequency $h\omega_0$, and $K$ is the truncation order, so that only harmonics with $|h| \le K$ are retained.
 \item Derivative Property (Frequency Shift): Differentiating a signal in the time domain translates into a state derivative plus an imaginary frequency offset in the harmonic domain:
\begin{equation}
    \left\langle \frac{dx}{dt} \right\rangle_h = \frac{d\langle x \rangle_h}{dt} + j h \omega_0 \langle x \rangle_h
\end{equation}
This offset creates diagonal matrices, such as the differentiation matrix $\mathcal{N} = \text{blkdiag}(j h \omega_0 I)$, that shift the system dynamics along the selected frequency.
 \item Linearisation: The non-linear differential equations governing the model $f(x, u)$ are expanded around a steady-state operating point using Taylor series or Jacobian linearisation to derive an LTP or LTI system. 
\end{itemize}

These steps transform the system from nonlinear and time-periodic to linear and time-invariant. Fourier series expansion and the derivative property are universal core transformations shared across all three frameworks. However, Toeplitz matrices are fundamental to HSS and DP, whereas DQsym replaces global Toeplitz matrices with a specialised 4-step phase-domain multiplication algorithm that involves transforming the multiplied variables to the abc frame, extracting Fourier coefficients from each signal, applying convolution, and packing the result into a complex sequence-dq signal for the DQsym solver~\cite {ALSARAYREH2026102719,alsarayreh_dynamic_2027}.

Multiplication of power converter state variables by non-linear switching functions in the time domain transforms into a discrete convolution across the harmonic domain
$  { \langle x y \rangle_h = \sum_{l} \langle x \rangle_{h-l} \langle y \rangle_l }$.
When expressed in matrix form, these convolutions are represented as block-Toeplitz matrices, which capture the coupling between harmonic orders.

Table~\ref{tab:dp_dqsym_hss_compact} compares the HD formulations in terms of state-matrix structure, sparsity and retained harmonic orders.

\begin{table}[h]
\caption{Comparison of DP, DQsym and HSS formulations}
\begin{center}
\scriptsize
\renewcommand{\arraystretch}{1.15}
\setlength{\tabcolsep}{2.5pt}
\begin{tabular}{p{0.16\columnwidth}p{0.245\columnwidth}p{0.245\columnwidth}p{0.245\columnwidth}}\toprule
Feature & DP & DQsym & HSS \\\midrule
State matrix form
  & Real-valued, with frequency-shift as $2\times2$ rotation blocks
  & Complex-valued sequence $dq$ $x_{dq}^{p,n,z}$
  & Block-Toeplitz minus frequency shift, $\mathbf{A}_{\mathrm{HSS}} = \mathcal{A} - \mathcal{N}$ \\
Retained harmonic order
  & $h \in \{0,\dots ,K\}$
  & $h \in \{0,\dots ,K\}$
  & contiguous $h \in \{-K,\dots ,K\}$ \\
State dimension
  & $(2\,K+1)\,N_{\mathrm{base}}$ (real-split)
  & $N_{\mathrm{solver}}$, each state transformed into complex dq, independent of K
  & $(2\,K+1)\,N_{\mathrm{base}}$ \\
Structure
  & Banded, with real $2\times2$ rotation blocks
  & Diagonal and decoupled per sequence and harmonic
  & Banded block-Toeplitz\\\botrule
\end{tabular}
\label{tab:dp_dqsym_hss_compact}
\end{center}\vspace*{-18pt}
\end{table}

\newcommand{\s}{\scriptstyle \Sigma}   
\renewcommand{\dd}{\scriptstyle \Delta}   
\newcommand{\su}{^\mathrm{u}}
\newcommand{\sll}{^\ell}
\newcommand{\jp}{\varphi}
\newcommand{\pluseq}{\mathrel{+}=}
\section{Benchmark Model of Modular Multilevel Converter (MMC)}
\label{sec:model}

\subsection{Open-loop MMC Model in the $abc$ Frame}
\label{sec:model_abc}

\begin{figure}[h]
    \centering \includegraphics[width=0.8\columnwidth]{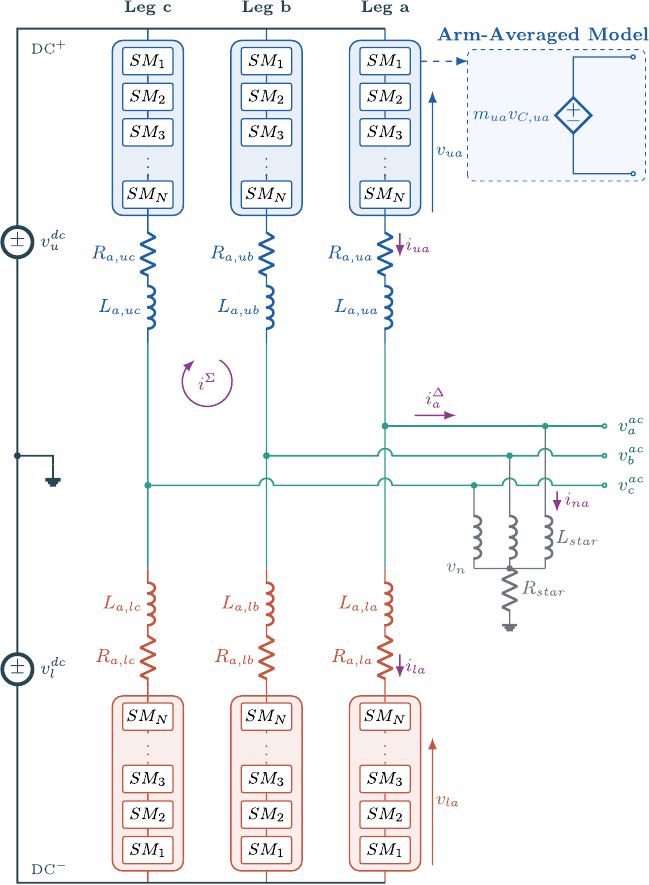}
    \caption{Three-phase MMC with half-bridge submodules: upper (blue) and lower (red) arms, split DC link, AC grid with star-point grounding, and showing the used arm-averaged model.}
    \label{fig:MMCcct}
\end{figure}

An MMC phase leg consists of an upper and a lower arm connected between the DC bus and the AC terminal, as in Fig.~\ref{fig:MMCcct}. Each arm contains a series inductor $L_{arm}$, a resistance $R_{arm}$ and an equivalent capacitor $C_{arm}=C_{SM}/N_{sm}$ representing the series connection of the submodule capacitors. The arm-averaged representation is used, with each arm described by its aggregated capacitor voltage and a controlled voltage source defined by its insertion index, which captures
the main electrical behaviour without modelling individual submodules. The model is open-loop with no controllers included, and the insertion indices are treated as inputs.  The AC side is represented by an ideal voltage source $\vec{v}^{g}$ at the converter terminals, and a star-point grounding reactor,
$R_{star}$-$L_{star}$ branch, is connected at the AC terminals.  The converter DC side is held constant at $v_{dc}^{u}$ and $v_{dc}^{l}$
with respect to ground.

The modulated arm voltages are, for $\jp\in\{a,b,c\}$,
\begin{equation}
v_{m\jp}^{u} = m_{\jp}^{u} v_{C\jp}^{u}, \qquad
v_{m\jp}^{l} = m_{\jp}^{l} v_{C\jp}^{l},
\label{modinabs}
\end{equation}
where $m_\jp^{u}$ and $m_\jp^{l}$ are the upper and lower insertion indices. The upper arm current $i_\jp^{u}$ flows from the positive pole to the AC terminal, and the lower arm current $i_\jp^{l}$ from the AC
terminal to the negative pole. Three-phase quantities are collected in vectors,  $\vec{i}^{u}=[i_a^{u}\; i_b^{u}\; i_c^{u}]^{T}$, $\vec{1} \triangleq [1\; 1\; 1]^{T}$, and $\odot$ denotes element-wise
multiplication. Applying Kirchhoff's voltage law in each arm, with the neutral voltage set to $v_n=0$, gives the averaged nonlinear model:
\begin{subequations}\label{eq:mmcdy}
\begin{align}
L_{arm} \frac{d \vec{i}^{u}}{dt} &= v_{dc}^{u}\vec{1} - \vec{v}^{g} - R_{arm}\vec{i}^{u} - \vec{m}^{u}\odot\vec{v}_{C}^{u}, \label{eq:mmcdy1} \\
L_{arm} \frac{d \vec{i}^{l}}{dt} &= -v_{dc}^{l}\vec{1} + \vec{v}^{g} - R_{arm}\vec{i}^{l} - \vec{m}^{l}\odot\vec{v}_{C}^{l}, \label{eq:mmcdy2} \\
C_{arm}\frac{d \vec{v}_{C}^{u}}{dt} &= \vec{m}^{u}\odot \vec{i}^{u}, \label{eq:mmcdy3}\\
C_{arm}\frac{d \vec{v}_{C}^{l}}{dt} &= \vec{m}^{l}\odot \vec{i}^{l}, \label{eq:mmcdy3b}\\
L_{star}\frac{d \vec{i}_{n}}{dt} &= \vec{v}^{g} - R_{star}\,\vec{1}\vec{1}^{T}\vec{i}_{n}. \label{eq:mmcdy4star}
\end{align}
\end{subequations}
With $v_n=0$ the three phases are decoupled, and $\vec{i}_n$ does not
influence the arm quantities.

For interpretation, the sum and difference variables:
\begin{equation}
\begin{aligned}
i_{\jp}^{\dd} &= i_{\jp}^{u} - i_{\jp}^{l}, &
i_{\jp}^{\s} &= \tfrac{1}{2}(i_{\jp}^{u} + i_{\jp}^{l}), \\
v_{C\jp}^{\dd} &= \tfrac{1}{2}(v_{C\jp}^{u} - v_{C\jp}^{l}), &
v_{C\jp}^{\s} &= \tfrac{1}{2}(v_{C\jp}^{u} + v_{C\jp}^{l}),
\end{aligned}
\label{eq:sumdiff}
\end{equation}
are used, where $i_\jp^{\dd}$ is the AC output current and
$i_\jp^{\s}$ the circulating current. They are obtained from the arm
quantities by a linear transformation and are used as outputs only. The models below are formulated in terms of upper- and lower-arm variables.

The AC source is a balanced positive-sequence voltage,
$\vec{v}^{g}(t)=V_g\,\vec{c}(\omega_0t)$
where $\vec{c}(\theta)=[\cos\theta\;\;\cos(\theta-\tfrac{2\pi}{3})\;\;
\cos(\theta+\tfrac{2\pi}{3})]^T$. The insertion indices are set in open
loop with a sinusoidal reference and no harmonic injection,
\begin{equation}
\vec{m}^{u}=\tfrac12\big(\vec{1}-\hat{m}\,\vec{c}(\omega_0t+\vartheta)\big),\qquad
\vec{m}^{l}=\tfrac12\big(\vec{1}+\hat{m}\,\vec{c}(\omega_0t+\vartheta)\big),
\label{eq:mod_index}
\end{equation}
and the pole potentials are $v_{dc}^{u}=V_{DC}/2$ and
$v_{dc}^{l}=-V_{DC}/2$. The parameters are listed in
Table~\ref{table:mmc_data}.

\begin{table}[!t]
\renewcommand{\arraystretch}{1.1}
\caption{Benchmark MMC Data}
\label{table:mmc_data}
\centering
\scriptsize
\begin{tabular}{lc}
\hline
\textbf{Parameter \& Description} & \textbf{Value} \\ \hline
$V_{DC}$: Pole-to-pole DC voltage & 640 kV \\ \hline
$R_{arm}$, $L_{arm}$: Arm resistance, inductance & 1.07 $\Omega$, 50 mH \\ \hline
$C_{SM}$, $N_{sm}$: Submodule capacitance, submodules per arm & 10 mF, 400 \\ \hline
$C_{arm}$: Equivalent arm capacitance & 25 $\mu$F \\ \hline
$R_{star}$, $L_{star}$: Star-point reactor resistance, inductance & 10 $\Omega$, 5 H \\ \hline
$V_g$: AC voltage (phase, peak) & 271.9 kV \\ \hline
$\hat{m}$, $\vartheta$: Modulation amplitude, angle & 0.85, $-2^\circ$ \\ \hline
$\omega_0$: Fundamental frequency & $2\pi \cdot 50$ rad/s \\ \hline
\end{tabular}
\end{table}

\begin{figure}[h]
    \centering
    \includegraphics[width=1\linewidth]{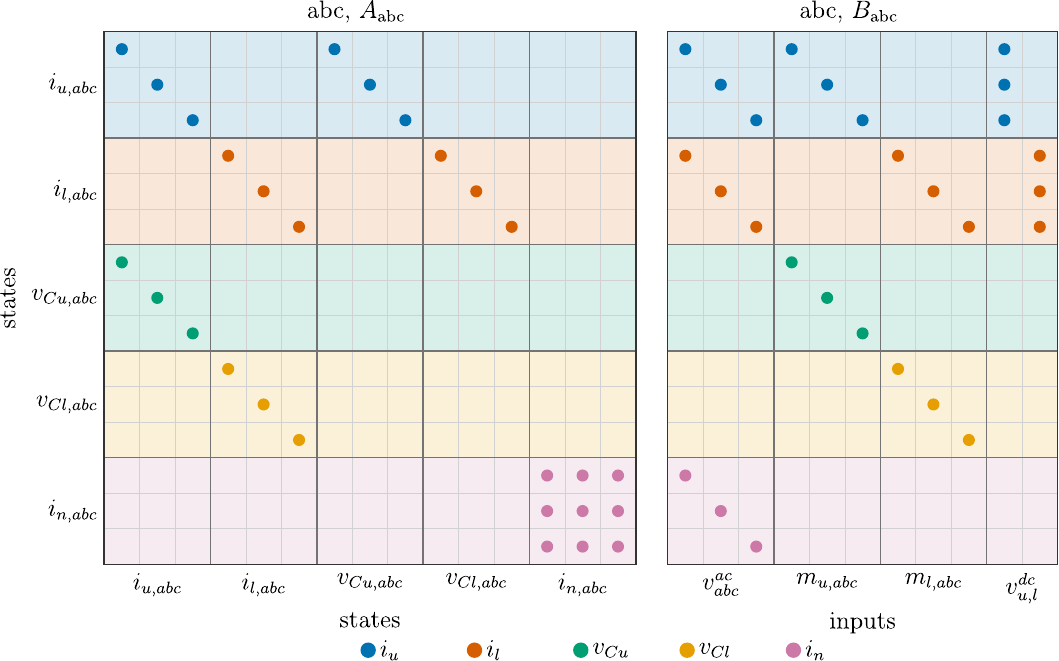}
    \caption{Sparsity pattern of the linearised open-loop abc arm-averaged MMC model. Each cell is one signal of phase $a$, $b$, or $c$. Dots are nonzero entries.}
    \label{fig:abcsparsity}
\end{figure}

Collecting \eqref{eq:mmcdy}, the state and input vectors are:
\begin{equation}
\begin{aligned}
\vec{x}&=\big[ \vec{i}^{u\,T}\;\;\vec{i}^{l\,T}\;\;\vec{v}_C^{u\,T}\;\;
\vec{v}_C^{l\,T}\;\;\vec{i}_n^{T}\big]^T\in\mathbb{R}^{15},\\
\vec{u}&=\big[\vec{v}^{g\,T}\;\;\vec{m}^{u\,T}\;\;\vec{m}^{l\,T}\;\;
v_{dc}^{u}\;\;v_{dc}^{l}\big]^T\in\mathbb{R}^{11}.
\end{aligned}
\label{eq:ss_xu}
\end{equation}
Every term in \eqref{eq:mmcdy} is either linear in the states and inputs or a product of an insertion index with a state, so the model is bilinear,
\begin{equation}
\dot{\vec{x}}=\vec{f}(\vec{x},\vec{u})=\mathbf{A}(\vec{u})\,\vec{x}+\mathbf{B}_0\,\vec{u},
\label{eq:ss_full}
\end{equation}
where $\mathbf{B}_0$ is constant and $\mathbf{A}(\vec{u})$ contains the
circuit parameters and the insertion indices. Since the insertion indices \eqref{eq:mod_index} are $T$-periodic with $T=2\pi/\omega_0$, the steady state is periodic, $\vec{x}_0(t)=\vec{x}_0(t+T)$. Linearising around $(\vec{x}_0(t),\vec{u}_0(t))$ gives the LTP system:
\begin{equation}
\begin{aligned}
\Delta\dot{\vec{x}}&=\mathbf{A}(t)\,\Delta\vec{x}+\mathbf{B}(t)\,\Delta\vec{u},\\
\mathbf{A}(t)&=\left.\frac{\partial\vec{f}}{\partial\vec{x}}\right|_{0}=\mathbf{A}\big(\vec{u}_0(t)\big),\qquad
\mathbf{B}(t)=\left.\frac{\partial\vec{f}}{\partial\vec{u}}\right|_{0}.
\end{aligned}
\label{eq:ss_lin}
\end{equation}
Because \eqref{eq:ss_full} is bilinear, the state Jacobian is simply
$\mathbf{A}(\vec{u})$ evaluated along the steady-state inputs, and the
same symbol $\mathbf{A}$ is used for it. In open loop, $\mathbf{A}(t)$
depends only on the insertion indices
$\vec{m}^{u}_0(t)$, $\vec{m}^{l}_0(t)$, whereas $\mathbf{B}(t)$
depends on the steady-state capacitor voltages and arm currents
through the columns associated with $\Delta\vec{m}^{u}$ and
$\Delta\vec{m}^{l}$.
Fig.~\ref{fig:abcsparsity} shows the sparsity pattern of the linearised open-loop abc arm-averaged MMC model, with $A_\text{abc}$ ($15\times15$) and $B_\text{abc}$ ($15\times11$) each have 27 nonzeros.


\subsection{MMC Model in harmonic domain formulation}
\label{sec:formulations}
 
This subsection derives the three harmonic-domain models of the benchmark
MMC. 
Linearising the open-loop state-space model in \eqref{eq:ss_full} around the periodic steady state and expanding the result in Fourier series gives the HSS model.
Applying generalised averaging to the nonlinear model instead gives
the DP model, and transforming each harmonic of the DP model to
sequence dq components gives the DQsym model. All three use the same
model, parameters and variables.

\subsubsection{HSS Model}
\label{sec:hss}

The HSS model is obtained from the linearised model
\eqref{eq:ss_lin} by representing its time-periodic matrices in the
harmonic domain. The periodic matrices are expanded as
$\mathbf{A}(t)=\sum_{h}\mathbf{A}_h e^{jh\omega_0t}$ and
$\mathbf{B}(t)=\sum_{h}\mathbf{B}_h e^{jh\omega_0t}$, with
${\mathbf{A}_h=\frac{1}{T}\int_0^T\mathbf{A}(t)e^{-jh\omega_0t}dt}$, the same expansion can be done to $\mathbf{B}_h$, and the signals as
$\Delta\vec{x}(t)=\sum_h\Delta\vec{X}_h\,e^{(s+jh\omega_0)t}$. Truncating to $|h|\le K$ and stacking
$\mathcal{X}=[\Delta\vec{X}_{-K}^T\;\cdots\;\Delta\vec{X}_{K}^T]^T$, likewise $\mathcal{U}$, gives the linear time-invariant HSS model:
\begin{equation}
s\,\mathcal{X}=\big(\mathcal{A}-\mathcal{N}\big)\mathcal{X}+\mathcal{B}\,\mathcal{U},
\label{eq:hss}
\end{equation}
\begin{equation}
\mathcal{A}=\begin{bmatrix}
\mathbf{A}_0 & \mathbf{A}_{-1} & \cdots & \mathbf{A}_{-2K}\\
\mathbf{A}_1 & \mathbf{A}_0 & \cdots & \mathbf{A}_{-2K+1}\\
\vdots & & \ddots & \vdots\\
\mathbf{A}_{2K} & \mathbf{A}_{2K-1} & \cdots & \mathbf{A}_0
\end{bmatrix},
\label{eq:toeplitz}
\end{equation}
where $\mathcal{N}=\mathrm{diag}(-K,\dots,K)\otimes j\omega_0\mathbf{I}_{15}$
and $\mathcal{B}$ is the block-Toeplitz matrix built from
$\mathbf{B}_h$ in the same way as \eqref{eq:toeplitz}.
For an MMC model with $N_{\mathrm{base}} = 15$ states and $M_{\mathrm{base}} = 11$ inputs, truncating at $K = 4$ yields an HSS model with $15(2K+1) = 135$ complex states and $11(2K+1) = 99$ complex inputs. 

\begin{figure}[h]
    \centering
    \includegraphics[width=1\columnwidth]{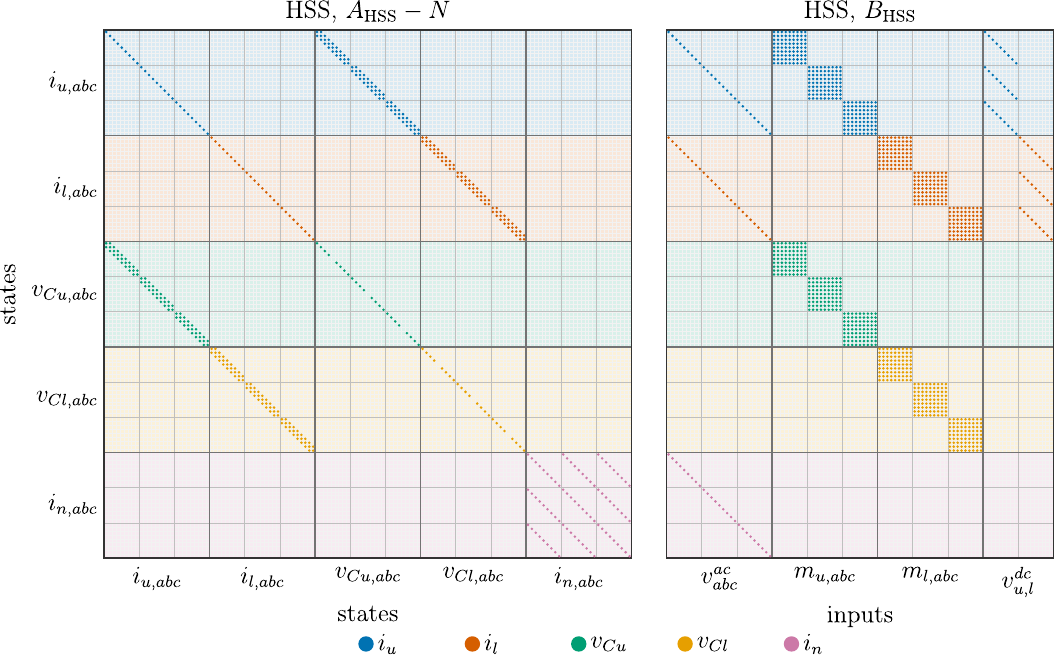}
    \caption{Sparsity pattern of the HSS model for $K=4$: $A_{\mathrm{HSS}}-N$ and $B_{\mathrm{HSS}}$. Dots are nonzero entries.}
\label{fig:sparsity_hss}
\end{figure}

Because \eqref{eq:mod_index} contains only the DC and fundamental
components, only $\mathbf{A}_0$ and $\mathbf{A}_{\pm1}$ are nonzero in \eqref{eq:toeplitz},
and each harmonic of the arm
quantities couples only to its neighbours $h\pm1$. 
Fig.~\ref{fig:sparsity_hss} shows the sparsity pattern of the HSS model for $K=4$, $\mathcal{A}_{\mathrm{HSS}}-\mathcal{N}$ has $135\times135$, 483 nonzeros, and $\mathcal{B}_{\mathrm{HSS}}$ has $135\times99$, 1107 nonzeros. Each signal block spans the $2K+1$ contiguous harmonic components $h \in \{-K, \dots  ,K\} $. The banded blocks in $\mathcal{A}_{\mathrm{HSS}}$ are the Toeplitz coupling through $\vec{m}$, and the full blocks in $\mathcal{B}_{\mathrm{HSS}}$ come from the harmonic content of $v_C(t)$ and $i(t)$.
\subsubsection{DP Model}
\label{sec:dp}

 
Whereas HSS linearises first and then expands in Fourier series, DP first transforms the nonlinear model \eqref{eq:ss_full} and then linearises it.

The $h$-th dynamic phasor of $x$ is
$\langle x\rangle_h(t)=\frac{1}{T}\int_{t-T}^{t}x(\tau)e^{-jh\omega_0\tau}d\tau$,
with the properties
\begin{equation}
\frac{d\langle x\rangle_h}{dt}=\Big\langle\frac{dx}{dt}\Big\rangle_h-jh\omega_0\langle x\rangle_h,\quad
\langle \vec{x}\odot\vec{y}\rangle_h=\sum_{l}\langle\vec{x}\rangle_{h-l}\odot\langle\vec{y}\rangle_{l}.
\label{eq:dpprop}
\end{equation}

Applying \eqref{eq:dpprop} to \eqref{eq:mmcdy} gives, for $0\le h\le K$
and with $\mathcal{D}_h=\tfrac{d}{dt}+jh\omega_0$,
\begin{subequations}\label{eq:mmcdp}
\begin{align}
L_{arm}\mathcal{D}_h\langle\vec{i}^{u}\rangle_h
&=\langle v_{dc}^{u}\rangle_h\vec{1}-\langle\vec{v}^{g}\rangle_h-R_{arm}\langle\vec{i}^{u}\rangle_h
-\langle\vec{v}_{m}^{u}\rangle_h,\\
L_{arm}\mathcal{D}_h\langle\vec{i}^{l}\rangle_h
&=-\langle v_{dc}^{l}\rangle_h\vec{1}+\langle\vec{v}^{g}\rangle_h-R_{arm}\langle\vec{i}^{l}\rangle_h
-\langle\vec{v}_{m}^{l}\rangle_h,\\
C_{arm}\mathcal{D}_h\langle\vec{v}^{u}_C\rangle_h
&=\textstyle\sum_{l}\langle\vec{m}^{u}\rangle_{h-l}\odot\langle\vec{i}^{u}\rangle_l,\\
C_{arm}\mathcal{D}_h\langle\vec{v}^{l}_C\rangle_h
&=\textstyle\sum_{l}\langle\vec{m}^{l}\rangle_{h-l}\odot\langle\vec{i}^{l}\rangle_l,\\
L_{star}\mathcal{D}_h\langle\vec{i}_n\rangle_h
&=\langle\vec{v}^{g}\rangle_h-R_{star}\vec{1}\vec{1}^{T}\langle\vec{i}_n\rangle_h,
\end{align}
\end{subequations}
where $\langle\vec{v}_{m}^{u}\rangle_h$ and $\langle\vec{v}_{m}^{l}\rangle_h$ are the DPs
of the modulated arm voltages \eqref{modinabs},
\begin{equation}
\langle\vec{v}_{m}^{u}\rangle_h=\sum_{l}\langle\vec{m}^{u}\rangle_{h-l}\odot\langle\vec{v}_C^{u}\rangle_l,\qquad
\langle\vec{v}_{m}^{l}\rangle_h=\sum_{l}\langle\vec{m}^{l}\rangle_{h-l}\odot\langle\vec{v}_C^{l}\rangle_l,
\label{eq:vmdp}
\end{equation}
The modulation \eqref{eq:mod_index} has
$\langle\vec{m}^{u}\rangle_0=\langle\vec{m}^{l}\rangle_0=\tfrac12\vec{1}$ and
$\langle\vec{m}^{l}\rangle_1=-\langle\vec{m}^{u}\rangle_1=\vec{\mu}_{1}$, where
$\vec{\mu}_{1}=\tfrac{\hat{m}}{4}e^{j\vartheta}[1\;\,a^2\;\,a]^T$ and $a=e^{j2\pi/3}$.
The pole voltages enter only at $h=0$ and the AC source only at $h=1$.

A real-valued model is obtained by taking the real and imaginary parts
of each DP as states,
\begin{equation}
\mathbf{X}=\big[\langle\vec{x}\rangle_{0}^T\;\;
\Re\langle\vec{x}\rangle_{1}^T\;\;\Im\langle\vec{x}\rangle_{1}^T\;\cdots\;
\Re\langle\vec{x}\rangle_{K}^T\;\;\Im\langle\vec{x}\rangle_{K}^T\big]^T,
\label{eq:mmcdy4}
\end{equation}
with $\mathbf{U}$ built from $\vec{u}$ in the same way, which gives the
nonlinear time-invariant DP model. For $K=4$, the state-space formulation of the DP model has 135 real states and 99 real inputs. The DP model is inherently time-invariant, and at steady state it has a constant equilibrium $(\mathbf{X}_0,\mathbf{U}_0)$. Linearising around this operating point gives an LTI representation.

\begin{figure}[h]
    \centering
    \includegraphics[width=1\columnwidth]{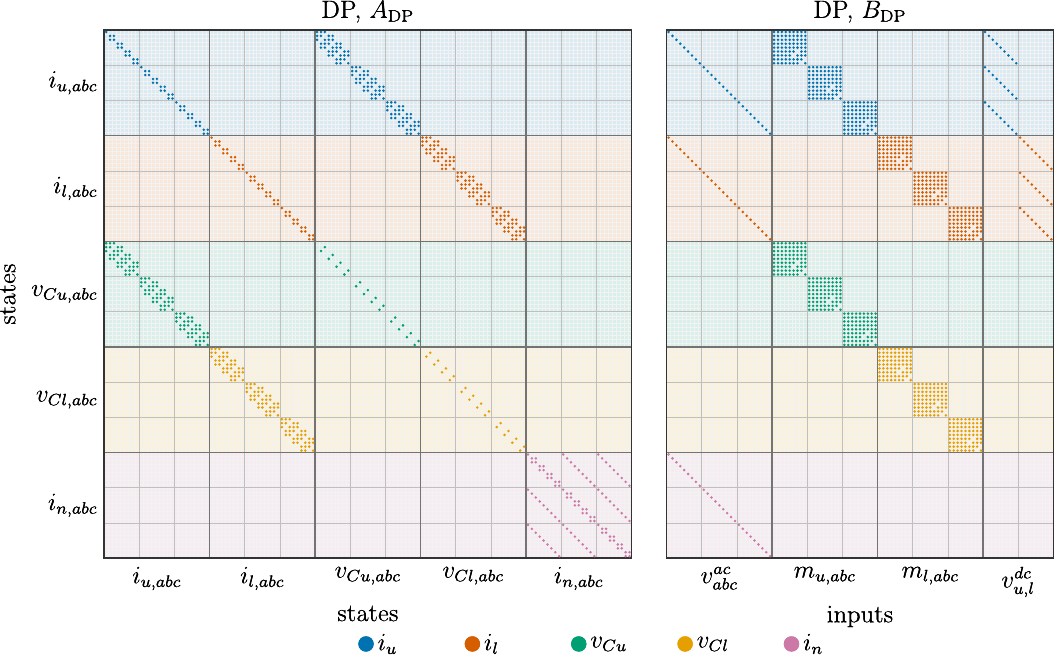}
    \caption{Sparsity pattern of the DP model for $K=4$. Dots are nonzero entries.}
\label{fig:sparsity_dp}
\end{figure}

Fig.~\ref{fig:sparsity_dp} shows the sparsity pattern of the DP model for $K=4$, $A_{\mathrm{DP}}$ has $135\times135$, 699 nonzeros and $B_{\mathrm{DP}}$ has $135\times99$, 1059 nonzeros. Each signal block holds $\langle x\rangle_0$ and the real and imaginary parts of $\langle x\rangle_1,\dots,\langle x\rangle_K$. The frequency shift $h\omega_0$ appears as $2\times2$ blocks in $A_{\mathrm{DP}}$, which makes it denser than $A_{\mathrm{HSS}}-\mathcal{N}$. 

\subsubsection{DQsym}
\begin{figure}[h]
    \centering
    \includegraphics[width=1\linewidth]{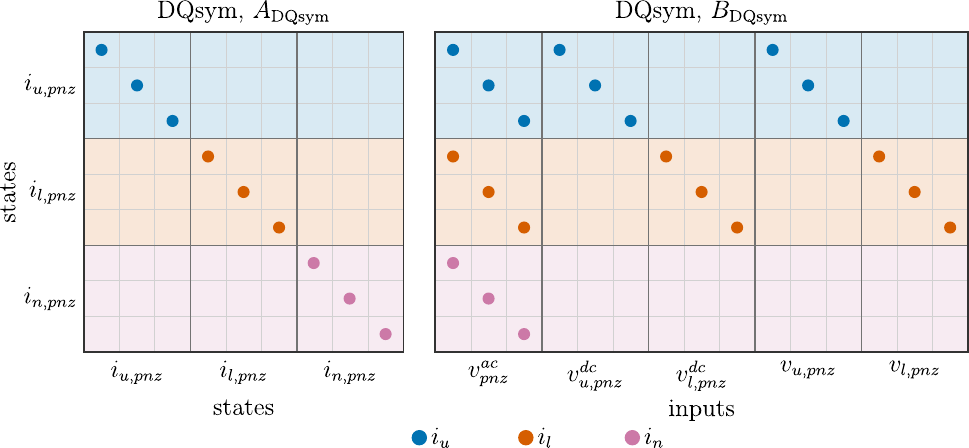}
    \caption{Sparsity patterns of the DQsym state matrix $A_{\mathrm{DQsym}}$ and $B_{\mathrm{DQsym}}$ in pnz coordinates. Dots are nonzero entries.}
    \label{fig:dqsym_sparsity}
\end{figure}

In the DQsym implementation, the model is split into two parts. 
The first is the arm current dynamics of \eqref{eq:mmcdy1}--\eqref{eq:mmcdy2},
\begin{equation}
\begin{aligned}
L_{arm} \tfrac{d \vec{i}^{u}}{dt} &= v_{dc}^{u}\vec{1} - \vec{v}^{g} - R_{arm}\vec{i}^{u} - \vec{v}_{m}^{u},\\
L_{arm} \tfrac{d \vec{i}^{l}}{dt} &= -v_{dc}^{l}\vec{1} + \vec{v}^{g} - R_{arm}\vec{i}^{l} - \vec{v}_{m}^{l},
\end{aligned}
\label{eq:core}
\end{equation}
which is linear and time-invariant. Because $L_{arm}$ and $R_{arm}$ are identical in all three phases, the symmetrical-component transformation
\begin{equation}
\vec{x}_{pnz} = \mathbf{T}\,\vec{x}_{abc},\qquad
\mathbf{T} = \frac{1}{3}
\begin{bmatrix}
1 & a & a^{2}\\
1 & a^{2} & a\\
1 & 1 & 1
\end{bmatrix},\quad a = e^{j2\pi/3},
\label{eq:fortescue}
\end{equation}
commutes with the arm impedance and decouples \eqref{eq:core} into independent positive-, negative-, and zero-sequence channels. Each sequence component is then expressed as a sum of complex dq phasors, one per retained harmonic order $h \in \{0, \dots, K\}$,
\begin{equation}
x_{s}(t) =  \sum_{h=0}^{K}(X_{s,h}^{d} + jX_{s,h}^{q}) e^{jh\omega_0 t},\quad s\in\{p,n,z\},
\label{eq:cdq}
\end{equation}
where $\omega_0$ is the fundamental angular frequency. Substituting \eqref{eq:cdq} into the sequence form of \eqref{eq:core} gives, for each harmonic order $h$, where $\mathbf{I}$ is the identity matrix,
\begin{equation}
\tfrac{d}{dt}\vec{X}_{h} = A_{h}\,\vec{X}_{h} + B_{h}\,\vec{U}_{h},\qquad
A_{h} = -\Big(\tfrac{R_{arm}}{L_{arm}} + jh\omega_0\Big)\mathbf{I},
\label{eq:dqsym_h}
\end{equation}
with states
$\vec{X}_{h}=[\vec{i}_{u,pnz,h}^{T}\;\vec{i}_{l,pnz,h}^{T}\;\vec{i}_{n,pnz,h}^{T}]^T$
and inputs
\begin{equation}
\vec{U}_{h}=\big[\vec{v}^{ac\,T}_{pnz,h}\;\;\vec{v}^{dc\,T}_{u,pnz,h}\;\;\vec{v}^{dc\,T}_{l,pnz,h}\;\;\vec{v}_{u,pnz,h}^{T}\;\;\vec{v}_{l,pnz,h}^{T}\big]^T ,
\label{eq:us}
\end{equation}
giving 9 states and 15 inputs forming the solver state-space matrix. The solver matrix $A_{DQsym}$  remains 9×9 for any $K$, and the harmonic order enters only through a diagonal frequency shift applied to each harmonic column of the signal.
As shown in Fig.~\ref{fig:dqsym_sparsity}, the solver state matrix $A_{\mathrm{DQsym}}$ is diagonal, and each row of the input matrix $B_{\mathrm{DQsym}}$ contains a single nonzero entry per input group, located in the same sequence channel as the state. Each sequence current is therefore driven only by the inputs of its own sequence, and no coupling between sequences or between harmonics appears in this part of the model.

The second part is the arm voltage calculation: the open-loop insertion
indices \eqref{eq:mod_index} are used directly to compute the capacitor
voltages \eqref{eq:mmcdy3}--\eqref{eq:mmcdy3b} and the modulated arm
voltages \eqref{modinabs},
\begin{equation}
\vec{v}_{m}^{u}=\vec{m}^{u}\odot\vec{v}_{C}^{u},\qquad
\vec{v}_{m}^{l}=\vec{m}^{l}\odot\vec{v}_{C}^{l},
\label{eq:w}
\end{equation}
which are transformed with \eqref{eq:fortescue} and fed to \eqref{eq:core} as the inputs $\vec{v}_{u,pnz}$ and $\vec{v}_{l,pnz}$. All harmonic coupling and nonlinearity are therefore confined to the arm voltage calculation, while the arm current dynamics remain diagonal and sparse.

\section{Results}
\label{sec:resluts}

All three HD models are evaluated on the same open-loop MMC of subsection~\ref{sec:model_abc}, with no circulating current control or energy control. The open-loop MMC has an AC side with a star-point reactor, as shown in Table~\ref{table:mmc_data}, and is used to investigate three forms of a single model. DP and HSS are truncated at $K = 4$ in their state matrices. DQsym uses the same truncation, $K = 4$, but applies it to the signals. Each signal carries harmonics $0 \to K$, and the truncation acts in the multiplication and integration blocks rather than in the solver matrix. The averaged arm model integrated directly in the
$abc$ frame (``abc ODE'') serves as the reference. All matrices in Fig.~\ref{fig:abcsparsity} to Fig.~\ref{fig:sparsity_dp} are shown in the same $abc$ harmonic basis, so the structural differences reflect only where the modulation-capacitor coupling is placed, not the choice of coordinates.

\subsection{Matrix structure and sparsity}

Table~\ref{tab:sparsity}
summarises the sizes and sparsity. The DQsym solver holds only the arm and neutral currents
($i_u$, $i_l$, $i_n$), giving 9 states per harmonic, 45 in total, instead of 135. The capacitor-voltage rows are evaluated outside the solver
by the multiplication and integration blocks. Therefore, $A_{\mathrm{DQsym}}$ contains only the per-harmonic diagonal blocks of the arm $RL$ branches, and it has no off-diagonal bands. In $A_{\mathrm{DP}}$ and
$A_{\mathrm{HSS}}-\mathcal{N}$, the modulation spectrum couples each arm
current to its capacitor voltage through banded Toeplitz blocks
($i_u\leftrightarrow v_{Cu}$, $i_l\leftrightarrow v_{Cl}$). $\mathbf{A}_{\mathrm{DP}}$
has the same pattern with each complex entry expanded into a $2\times 2$ real block.


The input matrices differ in the same way. In $B_{\mathrm{HSS}}$ and
$B_{\mathrm{DP}}$, the modulation inputs $\Vec{m}_u$, $\Vec{m}_l$ enter as dense
$(2K+1)\times(2K+1)$ blocks in both the current and the capacitor-voltage rows,
and linearising the products $\Vec{m}\,v_C$ and $\Vec{m}\,i$ uses the operating point of $v_C$ and $i$ at the periodic steady state. The DQsym solver instead takes the arm voltages as inputs,
which map onto the current rows harmonic by harmonic. Its input vector has 15 entries per harmonic, 75 in total, instead of 99.

\begin{table}[t]
\centering
\caption{State-space dimensions and sparsity for $K=4$.}
\label{tab:sparsity}
\small
\setlength{\tabcolsep}{3pt}
\begin{tabular}{lcccc}
\hline
 & DQsym & DQsym$^{\dagger}$ & DP & HSS \\
\hline
States / inputs         & 9 / 15   & 45 / 75   & 135 / 99   & 135 / 99 \\
nnz($A$) / nnz($B$)     & 9 / 21   & 45 / 105  & 699 / 1059 & 483 / 1107 \\
Density $A$ / $B$ (\%)  & 11.1 / 15.6 & 2.2 / 3.1 & 3.8 / 7.9 & 2.65 / 8.3 \\
Total nnz               & 30       & 150       & 1758       & 1590 \\
State type              & complex dq     & complex dq      & real       & complex \\
\hline
\end{tabular}\\[2pt]
\parbox{\columnwidth}{\footnotesize $^{\dagger}$Solver over all $K{+}1$ harmonics; excludes capacitor states and harmonic coupling, evaluated outside the solver.}
\end{table}

\subsection{Steady state}

\begin{figure}
    \centering
    \includegraphics[width=1\columnwidth]{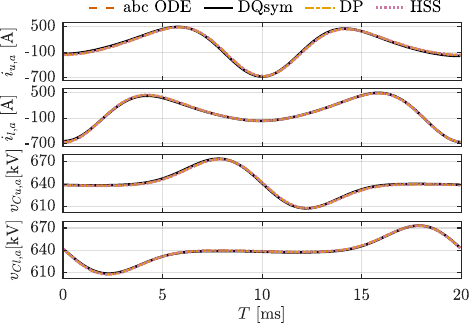}
    \caption{Periodic steady state of the open-loop MMC over one fundamental period: phase-$a$ upper and lower arm currents and arm capacitor voltages. Reference abc ODE compared with DQsym, DP, and HSS.}
    \label{fig:ss1cy}
\end{figure}

Fig.~\ref{fig:ss1cy} shows one fundamental period of the phase-$a$ upper and lower arm currents and arm capacitor voltages in periodic steady state. All three HD formulations overlap closely with the reference abc ODE model. The capacitor voltages ripple between about 608 and 673~kV around the 640~kV mean, and the upper-arm current reaches its minimum of about $-660$~A at $t=10$~ms in all models.

\begin{figure}
    \centering
    \includegraphics[width=1\columnwidth]{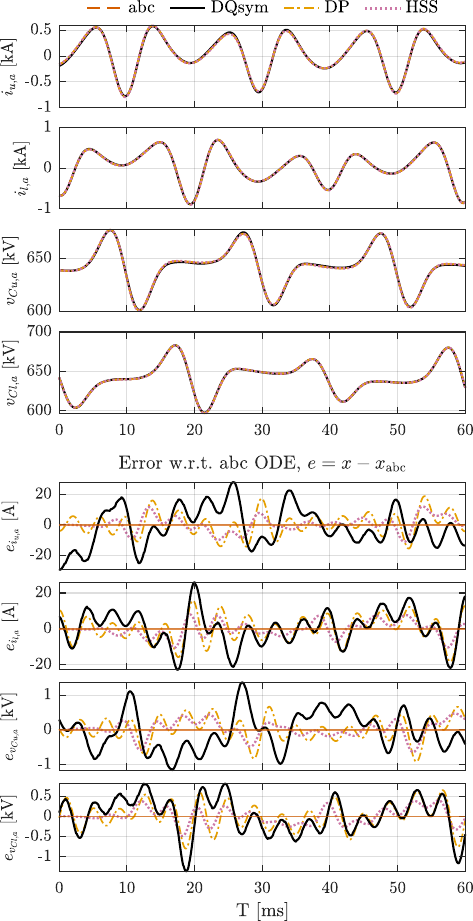}
    \caption{Response of the open-loop MMC over three fundamental periods: phase-$a$ upper and lower arm currents and arm capacitor voltages (top), and error of each model with respect to the reference abc ODE model (bottom).}
    \label{fig:ss4cy}
\end{figure}

Fig.~\ref{fig:ss4cy} extends the comparison to three fundamental periods and shows the error of each model with respect to the reference. The three formulations achieve comparable accuracy: for all three, the arm-current error remains below about 4\% of the peak arm current, and the capacitor-voltage error remains below about 0.2\% of the mean capacitor voltage. DQsym shows slightly larger peaks than DP and HSS (about 28~A versus 20~A in $i_{u,a}$), but its error is of the same order of magnitude and does not grow over the simulated interval.
This accuracy is obtained with the diagonal DQsym solver of Fig.~\ref{fig:dqsym_sparsity}, with the harmonic coupling evaluated outside it.

\subsection{Harmonic content}

\begin{figure}[t]
\centering
\includegraphics[width=\columnwidth]{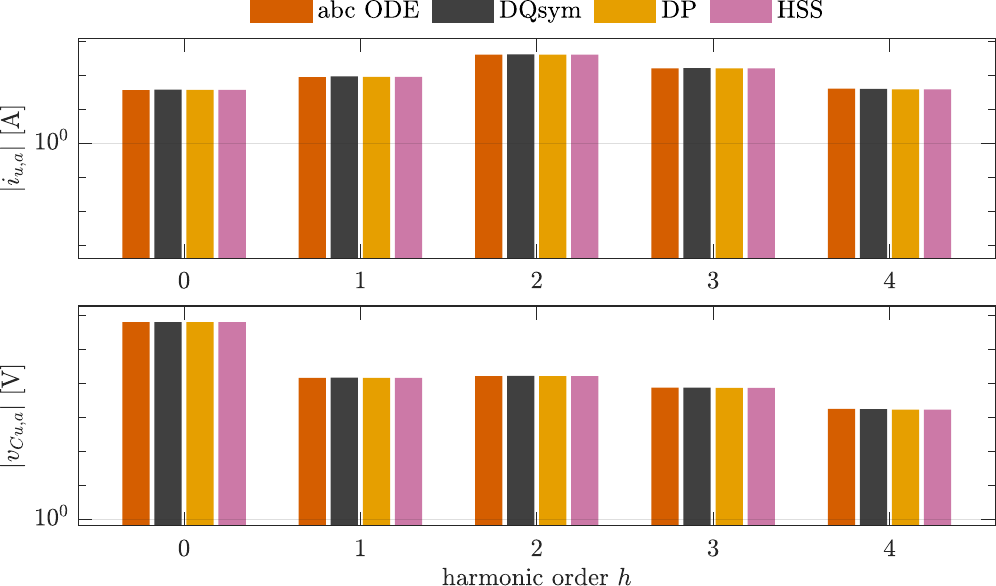}
\caption{Harmonic magnitudes of the phase-$a$ upper arm current and upper
capacitor voltage in periodic steady state (logarithmic scale).}
\label{fig:harmonics}
\end{figure}

Fig.~\ref{fig:harmonics} and Table~\ref{tab:harmonics} compare the harmonic
magnitudes of the phase-$a$ upper arm current and capacitor voltage in the
periodic steady state of Fig.~\ref{fig:ss1cy}, with the abc ODE as reference.
DP and HSS give identical spectra at every order.

All three formulations reproduce the reference closely, both for the dominant
components, such as the 2nd harmonic of the arm current (about 410~A) and the dc
capacitor voltage (about 640~kV), and for the smallest ones at $h=4$ (about 40~A
and 1.8~kV). The largest deviation is 4.2\% for DQsym in the 1st harmonic of
$i_{u,a}$, and 4.5\% for DP and HSS in the 4th harmonic of $v_{Cu,a}$.
DP and HSS are within 0.8\% of the
reference up to $h=3$ and deviate only at the truncation order $h=K=4$. DQsym deviates up to 4.2\% more uniformly
across the orders, but is closer to the reference at $h=4$, $-0.7\%$ and
$-1.7\%$, against $-4.5\%$ for DP and HSS.

\begin{table}[t]
\centering
\caption{Harmonic magnitudes of $|i_{u,a}|$ and $|v_{Cu,a}|$ for $h = 0,\dots,4$.}
\label{tab:harmonics}
\setlength{\tabcolsep}{4pt}
\begin{tabular}{llrrrrr}
\hline
 & & \multicolumn{5}{c}{Harmonic order $h$} \\
Quantity & Model & 0 & 1 & 2 & 3 & 4 \\
\hline
$|i_{u,a}|$ [A]
 & abc ODE  & 37.0 & 89.6 & 411.8 & 162.1 & 40.7 \\
 & DQsym    & 38.0 & 93.4 & 417.0 & 164.2 & 40.4 \\
 & DP / HSS & 37.3 & 90.2 & 411.7 & 161.3 & 38.9 \\
\hline
$|v_{Cu,a}|$ [kV]
 & abc ODE  & 639.6 & 14.70 & 16.47 & 7.52 & 1.79 \\
 & DQsym    & 640.0 & 14.90 & 16.69 & 7.61 & 1.76 \\
 & DP / HSS & 639.6 & 14.72 & 16.46 & 7.49 & 1.71 \\
\hline
\end{tabular}
\end{table}

\subsection{Eigenvalues}

DQsym evaluates $v_{arm}=\Vec{m}\,v_C$ and the capacitor integration outside
the solver. Writing this part as $\dot x_C = A_C x_C + B_C\, i$,
$v_{arm} = C_C x_C$, where $B_v$ maps the arm voltages into the solver and $C_{RL}$ extracts the arm currents, and closing it around the solver $(A_{RL},B_v,C_{RL})$ gives
\begin{equation}
A_{\mathrm{cl}} =
\begin{bmatrix} A_{RL} & B_v C_C \\ B_C C_{RL} & A_C \end{bmatrix},
\label{eq:Acl}
\end{equation}
where $A_{RL} \equiv A_\text{DQsym}$. $A_\text{cl}$ is similar to the linearised abc ODE model, with $B_vC_C$ and $B_CC_{RL}$ as the
modulation couplings. The loop has no algebraic path, and the Park and sequence
transformations act as a similarity transformation, so the eigenvalues of
$A_{\mathrm{cl}}$ can be compared directly with DP and HSS.

\begin{figure}
    \centering
    \includegraphics[width=\columnwidth]{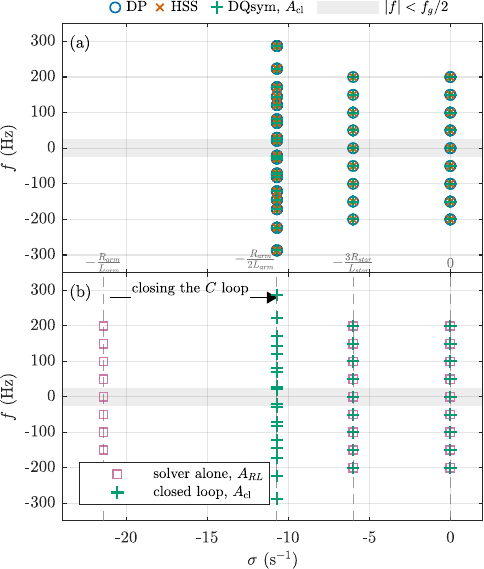}
    \caption{Eigenvalues of the open-loop MMC for $K=4$: (a) DP, HSS and DQsym
    ($A_{\mathrm{cl}}$); (b) DQsym solver alone and closed loop. The shaded band
    marks $|f|<f_g/2$.}
    \label{fig:eig}
\end{figure}

The eigenvalues of the three models are identical, as seen in Fig.~\ref{fig:eig}(a), with only three real parts, $\sigma = 0$, $-6$ and $-10.7\,\mathrm{s^{-1}}$,
and repeat at shifts of $\pm h f_g$ as these copies stem from the harmonic
formulation, so the distinct modes are those in $|f|<f_g/2$ with $f_g = \omega_0/2\pi$.

The solver alone, Fig.~\ref{fig:eig}(b), has real eigenvalues at
$-R_{arm}/L_{arm} = -21.4$, $-3R_{star}/L_{star} = -6$ and $0\,\mathrm{s^{-1}}$, and repeated at $\pm h f_g$.
The last two belong to the star branch, where the shared $R_{star}$ damps only
the zero sequence, and they do not involve the capacitors and are unchanged in the
closed loop. The arm pole moves to $-R_{arm}/(2L_{arm}) = -10.7\,\mathrm{s^{-1}}$
and becomes oscillatory, as the capacitors turn each first-order $RL$ arm into an
underdamped second-order one. All twelve arm modes take this value, showing that the arm resonance exists only after adding the capacitor dynamics.

\section{Conclusion}
\label{sec:conclusion}
This paper compared three harmonic-domain formulations of the same MMC model:
dynamic phasors (DP), harmonic state space (HSS) and DQsym. They differ only in
where the harmonic coupling between the modulation and the arm capacitor
voltages is placed. HSS and DP store this coupling in the state matrix, whereas
DQsym evaluates it outside the solver through multiplication and integration
blocks. Closing the DQsym solver with the capacitor dynamics recovers the same
linear time-periodic system, as confirmed by the identical eigenvalues of all
three formulations.

The formulations also differ in size and structure. For a truncation order $K=4$,
HSS and DP both have $135$ states, and their input matrices are dominated by
full $(2K{+}1)\times(2K{+}1)$ blocks, so the number of nonzeros grows
quadratically with $K$ ($1590$ for HSS and $1758$ for DP). The DQsym solver
contains only the $RL$ network in sequence coordinates. It is diagonal with 9 states (45 for $K = 4$), each a complex signal of width $K{+}1$, so it stays $9\times9$ for any $K$. The harmonic coupling is not removed but moved into the multiplication and integration blocks. All three
reproduce the steady-state harmonic content of the time-domain model
up to $h=4$. HSS is the natural choice for
analytical stability and impedance studies, DP for real-valued time-domain
simulation, and DQsym when modularity and a $K$-independent solver structure matter.

Because the solver alone does not contain the capacitor dynamics, stability
analysis with DQsym must linearise the solver together with its multiplication
and integration blocks, as done here. Future work will extend the comparison to
closed-loop operation with circulating-current suppression and to asymmetric
operating points, use the DQsym linearisation for small-signal stability
analysis, and present a systematic procedure for assembling the DQsym system
matrices.




\section{References}
\bibliographystyle{unsrt}
\bibliography{bib/references}

\end{document}